\documentclass[12pt]{article}
\usepackage{html}
\usepackage{epsf}
\usepackage{psfig}
\begin{document}
\title{MATTERS OF GRAVITY, The newsletter of the APS Topical Group on 
Gravitation}
\begin{center}
{ \Large {\bf MATTERS OF GRAVITY}}\\ 
\bigskip
\hrule
\medskip
{The newsletter of the Topical Group on Gravitation of the American Physical 
Society}\\
\medskip
{\bf Number 29 \hfill Winter 2007}
\end{center}
\begin{flushleft}
\tableofcontents
\vfill\eject
\section*{\noindent  Editor\hfill}
David Garfinkle\\
\smallskip
Department of Physics
Oakland University
Rochester, MI 48309\\
Phone: (248) 370-3411\\
Internet: 
\htmladdnormallink{\protect {\tt{garfinkl-at-oakland.edu}}}
{mailto:garfinkl@oakland.edu}\\
WWW: \htmladdnormallink
{\protect {\tt{http://www.oakland.edu/physics/physics\textunderscore people/faculty/Garfinkle.htm}}}
{http://www.oakland.edu/physics/physics_people/faculty/Garfinkle.htm}\\

\section*{\noindent  Associate Editor\hfill}
Greg Comer\\
\smallskip
Department of Physics and Center for Fluids at All Scales,\\
St. Louis University,
St. Louis, MO 63103\\
Phone: (314) 977-8432\\
Internet:
\htmladdnormallink{\protect {\tt{comergl-at-slu.edu}}}
{mailto:comergl@slu.edu}\\
WWW: \htmladdnormallink{\protect {\tt{http://www.slu.edu/colleges/AS/physics/profs/comer.html}}}
{http://www.slu.edu//colleges/AS/physics/profs/comer.html}\\
\bigskip
\hfill ISSN: 1527-3431

\begin{rawhtml}
<P>
<BR><HR><P>
\end{rawhtml}
\end{flushleft}
\pagebreak
\section*{Editorial}

The next newsletter is due September 1st.  This and all subsequent
issues will be available on the web at
\htmladdnormallink 
{\protect {\tt {http://www.oakland.edu/physics/Gravity.htm}}}
{http://www.oakland.edu/physics/Gravity.htm} 
All issues before number {\bf 28} are available at
\htmladdnormallink {\protect {\tt {http://www.phys.lsu.edu/mog}}}
{http://www.phys.lsu.edu/mog}

Any ideas for topics
that should be covered by the newsletter, should be emailed to me, or 
Greg Comer, or
the relevant correspondent.  Any comments/questions/complaints
about the newsletter should be emailed to me.

A hardcopy of the newsletter is distributed free of charge to the
members of the APS Topical Group on Gravitation upon request (the
default distribution form is via the web) to the secretary of the
Topical Group.  It is considered a lack of etiquette to ask me to mail
you hard copies of the newsletter unless you have exhausted all your
resources to get your copy otherwise.

\hfill David Garfinkle 

\bigbreak

\vspace{-0.8cm}
\parskip=0pt
\section*{Correspondents of Matters of Gravity}
\begin{itemize}
\setlength{\itemsep}{-5pt}
\setlength{\parsep}{0pt}
\item John Friedman and Kip Thorne: Relativistic Astrophysics,
\item Bei-Lok Hu: Quantum Cosmology and Related Topics
\item Gary Horowitz: Interface with Mathematical High Energy Physics and
String Theory
\item Beverly Berger: News from NSF
\item Richard Matzner: Numerical Relativity
\item Abhay Ashtekar and Ted Newman: Mathematical Relativity
\item Bernie Schutz: News From Europe
\item Lee Smolin: Quantum Gravity
\item Cliff Will: Confrontation of Theory with Experiment
\item Peter Bender: Space Experiments
\item Jens Gundlach: Laboratory Experiments
\item Warren Johnson: Resonant Mass Gravitational Wave Detectors
\item David Shoemaker: LIGO Project
\item Peter Saulson and Jorge Pullin: former editors, correspondents at large.
\end{itemize}
\section*{Topical Group in Gravitation (GGR) Authorities}
Chair: \'{E}anna Flanagan; Chair-Elect: 
Dieter Brill; Vice-Chair: David Garfinkle. 
Secretary-Treasurer: Vern Sandberg; Past Chair: Jorge Pullin;
Delegates:
Bei-Lok Hu, Sean Carroll,
Vicky Kalogera, Steve Penn,
Alessandra Buonanno, Bob Wagoner.
\parskip=10pt

\vfill
\eject

\section*{\centerline
{The View from the NSF}}
\addtocontents{toc}{\protect\medskip}
\addtocontents{toc}{\bf GGR News}
\addcontentsline{toc}{subsubsection}{
\it The View from the NSF, by Beverly Berger}
\parskip=3pt
\begin{center}
Beverly Berger, National Science Foundation
\htmladdnormallink{bberger-at-nsf.gov}
{mailto:bberger@nsf.gov}
\end{center}

My main objective in this note is to tell you about various websites with
information relevant NSF's role in the support of gravitational physics
research.

For those of you interested in NSF's budget, I recommend two sources of
information. The first is the American Institute of Physics FYI: Science
Policy News (see 
\htmladdnormallink
{\protect {\tt {http://www.aip.org/fyi/}}}
{http://www.aip.org/fyi/}
). You can subscribe to receive
alerts by email or view the archive.  For example, see
\htmladdnormallink
{\protect {\tt {http://www.aip.org/fyi/2006/141.html}}}
{http://www.aip.org/fyi/2006/141.html}
for some background on FY2007 budget
prospects.
See also
\htmladdnormallink
{\protect {\tt {http://www.aip.org/fyi/2007/014.html}}}
{http://www.aip.org/fyi/2007/014.html}
for the latest information.
The second source is NSF's Office of Legislative and Public
Affairs. Click on 
\htmladdnormallink
{\protect {\tt {http://www.nsf.gov/about/congress/index.jsp}}}
{http://www.nsf.gov/about/congress/index.jsp}
for the latest
information on the progress of the President's Budget Request through
Congress and on 
\htmladdnormallink
{\protect {\tt {http://www.nsf.gov/about/budget/}}}
{http://www.nsf.gov/about/budget/}
to see the Budget Request
for any Fiscal Year (the FY2008 Budget Request will appear in early
February) and, correspondingly, what was actually passed.

For those of you planning on submitting proposals or who have awards, an
excellent site for learning how to interact with Fastlane is the Fastlane
Demonstration Site
(
\htmladdnormallink
{\protect {\tt {http://www.fastlane.nsf.gov/jsp/homepage/demo\_site.html}}}
{http://www.fastlane.nsf.gov/jsp/homepage/demo_site.html}
). You can log in
as a fictitious PI and click on all possible buttons to see what happens.
Aside: THE PHYSICS DIVISION'S TARGET DATE FOR SUBMISSION IS THE LAST
WEDNESDAY IN SEPTEMBER.

To find out what is supported by the Gravitational Physics Program (or any
other NSF program), use the Award Search at 
\htmladdnormallink
{\protect {\tt {http://www.nsf.gov/awardsearch/}}}
{http://www.nsf.gov/awardsearch/}
.
For example you can perform a ``key word'' search to find all award abstracts
containing ``Einstein.'' If you scroll down and click ``active awards only,''
you will find 211 such awards. If you click on Program Information near the
top of the page, you can find all awards (past or present) supported by the
Gravitational Physics Program. Fill in the relevant Element Code:
Gravitational Experiment (1243), Gravitational Theory (1244), Support of
LIGO Research (1252), and LIGO Operations and Advanced R\&D (1293).

Finally, the main NSF website (
\htmladdnormallink
{\protect {\tt {http://www.nsf.gov}}}
{http://www.nsf.gov}
)
is the starting point to
search for funding opportunities outside Gravitational Physics. Clicking on
Math, Physical Sciences under Program Areas (top left column on the page)
will take you to the website of the Directorate of Mathematical and Physical
Sciences. Other Program Areas include Crosscutting (for Major Research
Instrumentation, CAREER, REU Sites), International (to see the programs of
the Office of International Science and Engineering), Computer, Info. Sci.,
Eng. (for the CISE Directorate), and Cyberinfrastructure (for the Office of
CyberInfrastructure).  In addition, the website 
\htmladdnormallink
{\protect {\tt {http://www.nsf.gov/funding/}}}
{http://www.nsf.gov/funding/}
allows searching and browsing for active solicitations.

\vfill\eject

\section*{\centerline
{GGR program at the APS meeting in Jacksonville}}
\addtocontents{toc}{\protect\medskip}
\addcontentsline{toc}{subsubsection}{
\it GGR program at the APS meeting in Jacksonville}
\parskip=3pt
\begin{center}
David Garfinkle, Oakland University
\htmladdnormallink{garfinkl-at-oakland.edu}
{mailto:garfinkl@oakland.edu}
\end{center}
We have an exciting GGR related program at the upcoming APS April meeting
in Jacksonville, FL.
Our chair-elect, Dieter Brill did a remarkable job of putting 
this program together.\\
 
Saturday April 14, 9 am (approx)\\
                        Plenary Talk\\
                First Results from Gravity Probe B\\
                Francis Everitt, Stanford University\\

B6: Saturday April 14, 10:45\\
                Binary Black Holes: Orbits, Mergers and Waveforms\\
                        Session Chair: Pedro Marronetti\\
Carlos Lousto: Spin-orbit interactions in black-hole binaries\\
Joan Centrella: Binary Black Hole Mergers (draft title)\\
Pablo Laguna Binary: Black holes and their echoes in the Universe\\

E6: Saturday, April 14, 15:30\\
                        GGR Prize Session\\
                Session Chair: Richard Isaacson\\
Ronald W. P. Drever: TBA\\
Rainer Weiss: The current state of LIGO\\
              and the plans for the near term future (tentative)\\
Gabriela Gonzalez: TBA\\

H4: Sunday April 15 8:30\\
                Gravity Probe B (preliminary title) (joint with GPMFC)\\
                        Session Chair: Clifford Will\\
John P. Turneaure: The Gravity Probe B Science Instrument\\
Bradford W. Parkinson: Gravity Probe B \& Innovative Space Engineering\\
George M. Keiser: Gravity Probe B Data Analysis Challenges, Insights \& 
Results\\

K4: Sunday, April 15, 13:15\\
                        Classical and semi-classical gravity\\
                           Session Chair: James Isenberg\\
Robert Wald: Present status of quantum field theory in curved spacetime\\
James Isenberg: Black Hole Rigidity\\
Greg Galloway: On the Topology of Higher Dimensional Black Holes\\

M4: Sunday, April 15, 15:15\\
                History of General Relativity (joint with FHP)\\
                        Session Chair: K. Wali\\
Daniel Kennefick: Traveling at the Speed of Thought --\\
                  Proving the Existence of Gravitational Waves\\
Richard Isaacson: Development of LIGO -- a view from Washington\\
Ted Newman: Survey of the Developments of GR since 1915 - present.\\

T5: Monday, April 16, 13:30\\
                Gravitational Wave Astrophysics with LISA (joint with DAP)\\
                        Session Chair: Joan Centrella\\
Craig Hogan: Gravitational Wave Backgrounds and Bursts\\
             from Terascale Phase Transitions and Cosmic Superstrings\\
Sterl Phinney: Gravitational waves as probes of galactic nuclei\\
               and accretion physics\\
Marta Volontieri: Coevolution of galaxies and massive black holes\\

Y4: Tuesday, April 17 13:30\\
                 Recent developments in quantum gravity\\
                        Session Chair: Jorge Pullin\\
Parampreet Singh: Recent advances in Loop Quantum Cosmology\\
Simone Speziale: Graviton propagator from loop quantum gravity\\
Max Niedermaier: The Asymptotic Safety Scenario in Quantum Gravity\\

Also on the program will be a focus session\\
   ``Hydrodynamics and Magnetohydrodynamics Coupled to General Relativity''\\
featuring an invited talk by John Hawley on his work modeling magnetized 
accretion about a black hole. The session will include talks on recent 
advances coupling fluids to gravity, including neutron stars and 
accretion. Those working in this area are encouraged to submit abstracts.

Other sessions at the APS meeting that may be of interest to gravitational 
physicists include (but are not limited to) the following:

Plenary Session Monday April 16\\
        String Theory, Branes, and if You Wish, the Anthropic Principle\\
                        Shamit Kachru, Stanford University\\

                        Cosmology After WMAP\\
                David Spergel, Princeton University\\

Sunday, April 15  8:30\\
        Precision Experiments and Tests of Fundamental Laws (GPMFC)\\
Blayne Heckel: CP violation and preferred frame tests using polarized 
electrons\\
David Reitze: Science with LIGO\\
Karl Van Bibber: Axions\\

Sunday, April 15  10:30\\
                        Compact Inspirals (DAP)\\
Chris Deloye: Inspirals and Gravitational Waves\\
Ingrid Stairs: The Double Pulsar\\
Danny Steeghs: White Dwarf Inspirals\\

Monday, April 16  10:45\\
        Few Body Computational Challenges for Large Scale Astrophysics 
(DCOMP)\\
Harald Pfeiffer: Binary black hole coalescence\\

Tuesday, April 17 10:30\\
                        Black Holes of All Sizes (DAP)\\
Phil Kaaret: Intermediate-Mass Black Holes\\
Avi Loeb: Supermassive Black Holes\\
John Miller: Stellar Black Holes\\

Tuesday, April 17\\
                        Gravity and Cosmology (DPF)\\
D. Huterer: The Accelerating Universe, Dark Energy, and Modified Gravity\\
D. Kapner: Experimental Results on Gravity at Short Distances\\
J.Santiago: Gravitation and Extra Dimensions\\

\vfill\eject

\section*{\centerline
{we hear that \dots}}
\addtocontents{toc}{\protect\medskip}
\addcontentsline{toc}{subsubsection}{
\it we hear that \dots , by David Garfinkle}
\parskip=3pt
\begin{center}
David Garfinkle, Oakland University
\htmladdnormallink{garfinkl-at-oakland.edu}
{mailto:garfinkl@oakland.edu}
\end{center}

Rainer Weiss and Ronald Drever are this year's winners of the APS {\it 
Einstein Prize} for Gravitational Physics.

Joseph Polchinski and Juan Maldecena are this year's winners of the APS
{\it Dannie Heineman Prize} for Mathematical Physics.

Gabriela Gonzalez is this year's winner of the APS {\it 
Edward A. Bouchet Award}. 

Ed Seidel has received the Sidney Fernbach award of the IEEE. 

Frederick Raab and Jennie Traschen have been elected APS Fellows.

Jorge Pullin has been elected a corresponding member of the 
{\it Mexican Academy of Sciences} and the {\it National Academy of Sciences 
of Argentina}

Hearty Congratulations!

\section*{\centerline
{100 years ago}}
\addtocontents{toc}{\protect\medskip}
\addcontentsline{toc}{subsubsection}{
\it 100 years ago, by David Garfinkle}
\parskip=3pt
\begin{center}
David Garfinkle, Oakland University
\htmladdnormallink{garfinkl-at-oakland.edu}
{mailto:garfinkl@oakland.edu}
\end{center}
Einstein formulated the equivalence principle in ``On the relativity
principle and the conclusions drawn from it'' {\it Jarbuch der 
Radioactivitaet und Elektronik} {\bf 4} (1907)

\vfill\eject

\section*{\centerline
{The Double Pulsar -- A unique gravity lab}}
\addtocontents{toc}{\protect\medskip}
\addtocontents{toc}{\bf Research Briefs:}
\addcontentsline{toc}{subsubsection}{
\it The Double Pulsar, by Michael Kramer}
\parskip=3pt
\begin{center}
Michael Kramer, The University of Manchester 
\htmladdnormallink{michael.kramer-at-manchester.ac.uk}
{mailto:michael.kramer@manchester.ac.uk}
\end{center}

Almost a hundred years after Einstein formulated his theory of general
relativity (GR), efforts in testing GR and its concepts are still being made
by many colleagues around the world, using many different approaches.  To date
GR has passed all experimental and observational tests with flying colours,
but in light of recent progress in observational cosmology in particular, the
question of whether alternative theories of gravity need to be considered is
as topical as ever.

Many experiments are designed to achieve ever more stringent tests by either
increasing the precision of the tests or by testing different, new aspects.
Some of the most stringent tests are obtained by satellite experiments in the
solar system, providing exciting limits on the validity of GR and alternative
theories of gravity like tensor-scalar theories.  However, solar-system
experiments are made in the gravitational weak-field regime, while deviations
from GR may appear only in strong gravitational fields. It happens that nature
provides us with an almost perfect laboratory to test the strong-field regime
using binary radio pulsars.

While, strictly speaking, the binary pulsars move in the weak gravitational
field of a companion, they do provide precision tests of the strong-field
regime. This becomes clear when considering strong self-field effects which
are predicted by the majority of alternative theories. Such effects would, for
instance, clearly affect the pulsars' orbital motion, allowing us to search for
these effects and hence providing us with a unique precision strong-field test
of gravity.

Pulsars are highly magnetized rotating neutron stars and are unique and
versatile objects which can be used to study an extremely wide range of
physical and astrophysical problems.  Besides testing theories of gravity one
can study the Galaxy and the interstellar medium, stars, binary systems and
their evolution, plasma physics and solid state physics under extreme
conditions.  This wide range of applications is exemplified by the first ever
discovered double pulsar \cite{bdp+03,lbk+04}. This unique system allows us to
test many aspects of gravitational theories at the same time, representing a
truly unique laboratory for relativistic gravity. The experiment is
conceptually simple: Nature has provided us with two clocks attached to point
masses which fall in the gravitational potential of their companion. Measuring
the ticks of these clocks while they move through space-time allows us to
compare our observations with the predictions of various theories of gravity.

The double pulsar is a system of two visible radio pulsars with periods of
22.8 ms (PSR J0737$-$3039A, simply called ``A'' hereafter) and 2.8 s (PSR
J0737$-$3039B, simply called ``B'' hereafter), respectively. It
was discovered and is studied by a large collaboration involving colleagues
from Australia, Canada, India, Italy and USA. The double pulsar's short and
compact (orbital period of $P_b = 144$ min), slightly eccentric ($e=0.09$)
orbit makes the double pulsar the most extreme relativistic binary system ever
discovered, demonstrated by the system's remarkably high value of periastron
advance ($\dot{\omega}=16.8995\pm0.0007\deg$ yr$^{-1}$, i.e.~four times larger
than for the Hulse-Taylor pulsar!). Only four years after the discovery of the
system, most of its timing parameters are determined with a precision that
took several decades to achieve in the previously known best relativistic
binary pulsars \cite{ksm+06}.
For instance, we measure that the orbit is shrinking every day
by $7.42\pm0.09$ mm, which agrees with GR's prediction of an orbital decay due
to the emission of gravitational quadrupole waves within an uncertainty of
1\%.  Ultimately, the shrinkage leads to a coalescence of the two pulsars in
only $\sim 85$ Myr.  This boosts the hopes for detecting a merger of two
neutron stars with first-generation ground-based gravitational wave detectors
by a factor of several compared to previous estimates \cite{bdp+03,kkl+04}.
Moreover, the detection of a young companion B around an old millisecond
pulsar A confirms the evolution scenario proposed for the creation of recycled
millisecond pulsars.

The measured precession of the orbit and the decrease in orbital period of
$\dot P_{\rm b}= (1.25\pm0.2)\times 10^{-12}$ seconds per second are both
observed deviations from a pure Keplerian description of the orbit.  It is
important to note that we do not have to assume a particular theory of gravity
when measuring such relativistic corrections, called ``post-Keplerian'' (PK)
parameters. Instead, we can take the observational values and compare them
with predictions made by a theory of gravity to be tested. In the double
pulsar, as A has the faster pulse period, we can time A much more accurately
than B, allowing us to measure a total of five very precise PK corrections for
A's orbit.

The PK parameter, $\dot{\omega}$, is the easiest to measure. When
interpreting this advance of periastron in the framework of GR,
it provides an immediate measurement of the total mass of the system.  The PK
parameter $\gamma$ denotes the amplitude of delays in arrival times caused by
the varying effects of the gravitational redshift and time dilation (second
order Doppler) as the pulsars move in an elliptical orbit at varying distances
with varying speeds. As a result of the gravitational redshift, the pulsar
clocks slow down when they 'feel' the deeper gravitational potential of the
companion and speed up when they are further away.

As mentioned, the decay of the orbit due to gravitational wave damping is
observed as a change in orbital period, $\dot{P}_{\rm b}$. Two further PK
parameters, $r$ and $s$, are related to a Shapiro delay caused by the
curvature of space time near the companion. Their measurement is possible,
since -- quite amazingly! -- we observe the system almost completely edge-on.
Hence, at superior conjunction the pulses of A pass the surface of B in only
30,000 km distance, needing to travel an extra length of curved space-time and
adding about 100 microseconds to the travel time to Earth. Within GR, we can
interprete $s$ as the sine of the orbital inclination angle. With a
measurement of $\sin i \equiv s = 0.99974(-0.00039,+0.00016)$ ,
this is indeed very close to an edge-on geometry of $i=90$deg.

When trying to see whether these PK parameter measurements are in agreement
with the predictions of GR or any other theory of gravity, we use that for
point masses with negligible spin contributions the PK parameters in each
theory should only be functions of the a priori unknown neutron star masses
and the well measurable Keplerian parameters.  With the two masses as the only
free parameters, the measurement of three or more PK parameters
over-constrains the system, and thereby provides a test ground for theories of
gravity.  These tests can be illustrated in a very elegant way \cite{dt92}:
The unique relationship between the two masses of the system predicted by any
theory for each PK parameter can be drawn in a diagram showing the mass of A
on one axis and that of B on the other. We expect all curves to intersect in a
single point if the chosen theory is a valid description of the nature of this
system (see figure).

\begin{figure*}
\centerline{\psfig{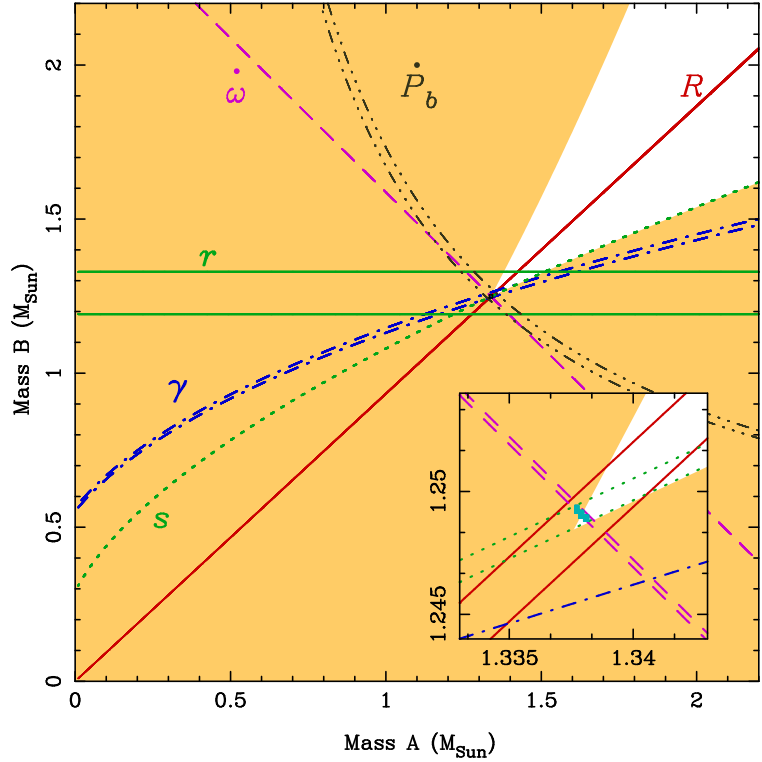}}
\caption{\label{fig:m1m2} `Mass--mass' diagram showing the
observational constraints on the masses of the neutron stars in the
double pulsar system J0737--3039.  The shaded regions are those that
are excluded by the Keplerian mass functions of the two
pulsars. Further constraints are shown as pairs of lines enclosing
permitted regions as given by the observed mass ratio, $R$, and the PK
parameters shown here as predicted by general relativity (see text).
Inset is an enlarged view of
the small square encompassing the intersection of these
constraints. See Kramer, Stairs, Manchester et al.~(2006) for details.}
\end{figure*}

Most importantly, the possibility to measure the orbit of both A and B
provides a new, qualitatively different constraint in such an analysis.
Indeed, with a measurement of the projected semi-major axes of the orbits of
both A and B, we obtain a precise measurement of the mass ratio simply from
Kepler's third law, via $R \equiv M_A/M_B = x_B/x_A$ where $M_A$ and $M_B$ are
the masses and $x_A$ and $x_B$ are the (projected) semi-major axes of the
orbits of both pulsars, respectively.  We can expect the mass ratio, $R$, to
follow this simple relationship to at least 1PN order.  In
particular, the $R$ value is not only theory-independent, but also independent
of strong-field (self-field) effects which is not the case for PK-parameters.
Therefore, any combination of masses derived from the PK-parameters {\em must}
be consistent with the mass ratio derived from Kepler's 3rd law. With five PK
parameters already available, this additional constraint makes the double
pulsar the most overdetermined system to date where the most relativistic
effects can be studied in the strong-field limit. The theory of GR passes this
new test at the record-breaking level of 0.05\% \cite{ksm+06}.

The precision of the measured timing system parameters increases continuously
with time as further and better observations are made.  Soon, we expect the
measurement of additional PK parameters, allowing more and new tests of
theories of gravity. Some of these parameters arise from a relativistic
deformation of the pulsar orbit and those which find their origin in
aberration effects and their interplay with geodetic precession.
In a few years, we will measure the decay of the orbit so accurately, that we
can put limits on alternative theories of gravity which should even surpass
the precision achieved in the solar system. On somewhat longer time scales, we
will even achieve a precision that will require us to consider post-Newtonian
terms that go beyond the currently used description of the PK parameters.
Indeed, we already achieve a level of precision in the $\dot \omega$
measurement where we expect corrections and contributions at the 2PN level.
One such effect involves the prediction by GR that, in contrast to Newtonian
physics, the neutron stars' spins affect their orbital motion via spin-orbit
coupling. This effect modifies the observed $\dot{\omega}$ by an amount that
depends on the pulsars' moment of inertia, so that a potential measurement of
this effect would allow the moment of inertia of a neutron star to be
determined for the very first time \cite{ds88,lbk+04}. We do not expect
this measurement to be easy, but we will certainly try!

With the measurement of already five PK parameters and the unique information
about the mass ratio, the double pulsar indeed provides a truly unique
test-bed for relativistic theories of gravity. Again, GR has passed these
new tests with flying colours. The precision of these tests and the nature of
the resulting constraints go beyond what has been possible with other systems
in the past.  However, we only just started to study and exploit the
relativistic phenomena that can be investigated in great detail in this
wonderful cosmic laboratory.


\vfill\eject

\section*{\centerline
{Theoretical Approaches to Cosmic Acceleration}}
\addtocontents{toc}{\protect\medskip}
\addcontentsline{toc}{subsubsection}{
\it Theoretical Approaches to Cosmic Acceleration, by Mark Trodden}
\parskip=3pt
\begin{center}
Mark Trodden, Syracuse University
\htmladdnormallink{trodden-at-physics.syr.edu}
{mailto:trodden@physics.syr.edu}
\end{center}

Less than a decade ago, observations of the lightcurves of type Ia supernovae first suggested that the expansion of the universe is accelerating. In the intervening years, a range of further observations~\cite{Knop03,Riess04} have provided
firm support for this result, to the extent that, even if we were to ignore the
supernova data entirely, the accelerating universe would remain unavoidable.

At the level of cosmological models, described by perfect fluids with phenomenological equations of state, the accelerating universe just requires new parameters to fit the known data. Augmenting the general Cold-Dark-Matter (CDM) cosmology with an extra fluid component, X, with present-day energy density $\rho_X$ and
constant equation of state parameter $w_X$, satisfying $p_X=w_x \rho_X$, the Friedmann equation becomes
\begin{equation}
H^2=\frac{1}{3M_p^2}\left[\rho_{\rm m}\left(\frac{a_0}{a}\right)^3 +\rho_X \left(\frac{a_0}{a}\right)^{3(1+w_X)}\right] -\frac{k}{a^2} \ .
\end{equation}

It is to this parametrization (with other parameters determining the initial spectrum of perturbations) that cosmological datasets are fit, perhaps the best-known being the WMAP data.

That such a small number of parameters can provide such a tremendous fit to the
evolution of the universe, including its large-scale structure, over its entire
history, is a triumph of modern cosmology comparable to the broad successes of the expansion, the discovery of the CMB and the agreement of the abundances of the light elements. However, such an approach, while remarkably useful, does not provide an {\it explanation} for the origin of cosmic acceleration. Indeed, the biggest impact of the accelerating universe is in its implications for fundamental physics.

Clearly, one possibility is that cosmic acceleration is due to the cosmological
constant (with $w_X=-1$). The cosmological constant problem itself -- why is the vacuum energy so much smaller than we expect from effective-field-theory considerations? -- requires a solution even in the absence of cosmic acceleration, and perhaps the final answer to this problem will yield a value appropriate to lead to late-time acceleration of the universe. Despite continuous theoretical pressure, the status of dynamical solutions to this conundrum has changed little since Weinberg's review article of 1988~\cite{Weinberg:1988cp}. Historically, this has led some researchers to consider an anthropic solution to the problem, although without a specific fundamental framework in which to investigate it.

However, in the context of string theory, the possibility of a {\it landscape},
containing at least $10^{100}$ discrete vacua with vacuum energy densities ranging up to the Planck scale, coupled with the mechanism of eternal inflation to populate the landscape, has recently led to a specific implementation of the anthropic argument.

While such a conclusion would seem to limit the testability of the proposal, one hope is that it might be possible for the statistics of the distribution of vacua \cite{Dine05,Garriga05, Hamed05, Barnard05,Tegmark05, Burgess05}, to allow statistical predictions for other observable quantities, such as the fundamental coupling constants. Should increasingly accurate cosmological observations reveal a dark energy equation of state not equal to -1, or evidence for temporal or spatial variation of the dark energy density, then we will know that a cosmological constant is not the answer and it will be harder to imagine anthropic arguments from the string landscape being the correct answer.

If the cosmological constant is not responsible for dark energy (because it is zero, or much smaller than the dark energy scale), then several possibilities have been suggested for a dynamical origin for cosmic acceleration.

The first of these - {\it dark energy}~\cite{Wetterich88,Ratra88,Caldwell:1997ii} - seeks to find an underlying microscopic description of the perfect fluid. The most popular approach to this new dynamical component of the cosmic energy budget is to invoke a new scalar field driving late-time inflation (but without the need for an end, as in the reheating that takes place after early universe inflation). In such a {\it quintessence} model, the instantaneous effective dark energy equation of state is
\begin{equation}
w_{\phi}=\frac{{\dot \phi}^2-2V(\phi)}{{\dot \phi}^2+2V(\phi)} \ .
\end{equation}

If one assumes that cosmic acceleration is due to such a field, with potential-dominated dynamics, then generally one finds that the scale of the potential ($V^{1/4}$) should be of order $10^{-3}$eV, and that the mass of the associated particle be of order the Hubble scale. These scales present obstacles to finding a sensible particle physics model of quintessence. One that does seem to work, with such technically natural parameter values, is if the quintessence field is realized as the pseudo-Nambu-Goldstone boson of some broken symmetry. In this case the unusually small values of parameters required are protected from quantum corrections by the symmetry.

One advantage of a subset of quintessence models is that they exhibit {\it tracking}. This means that there exist attractors of the dynamical system for which the scalar field tracks the equation of state of the background fluid. It can then be arranged that the field follows the evolution of the universe during radiation domination and then transitions to an accelerating attractor during matter domination. This allows a partial explanation of the coincidence problem, since acceleration is triggered by the onset of matter domination.

Another interesting suggestion has been that it may be possible to explain an accelerated universe by invoking the
effects of inhomogeneities on the expansion rate -- perturbations may induce an
effective energy-momentum tensor with a
nearly-constant magnitude. Kolb et. al. \cite{Kolb05} have considered sub-horizon higher order corrections to the backreaction, going up to sixth order in a gradient expansion, and suggest that higher order corrections are large enough for
the backreaction to generate dark energy like behavior. There have been a number of challenges to this idea (see e.g.~\cite{Ishibashi05}), but if a successful mechanism is found it would be an elegant and minimal explanation of acceleration.

A further possibility is that curvatures and length scales in the observable universe are only now reaching values at which an infrared modification of gravity
can make itself apparent by driving self-acceleration. This possibility turns out to be incredibly difficult to implement.

Although, within the context of General Relativity (GR), one doesn't think about it too often, the metric tensor contains, in principle, more degrees of freedom than the usual spin-2 {\it graviton}. However, the Einstein-Hilbert action results in second-order equations of motion that constrain away the scalars and the
vectors, so that they are non-propagating. But this is not the case if one departs from the Einstein-Hilbert form for the action. When using any modified action (and the usual variational principle) one inevitably frees up some of the additional degrees of freedom. In fact, this can be a good thing, in that the dynamics of these new degrees of freedom may be precisely what one needs to drive the accelerated expansion of the universe. In many situations though, there is a price to pay.

The problems may be of several different kinds. First, there is the possibility
that along with the desired deviations from GR on cosmological scales, one may also find similar deviations on solar system scales, at which GR is rather well-tested. Second is the possibility that the newly-activated degrees of freedom may be badly behaved in one way or another; either having the wrong sign kinetic terms (ghosts), and hence being unstable, or leading to superluminal propagation,
which may lead to other problems.
These constraints are surprisingly restrictive when one tries to create viable modified gravity models yielding cosmic acceleration.

As an example, one simple way to modify GR is to replace the Einstein-Hilbert Lagrangian density by a general function $f(R)$ of the Ricci scalar $R$. For appropriate choices of the function $f(R)$ it is then possible to obtain late-time cosmic acceleration without the need for dark energy~\cite{Carroll:2003wy}. However, evading bounds from precision solar-system tests of gravity turns out to be a much trickier matter, since such simple models are equivalent to a Brans-Dicke
theory with $\omega=0$ in the approximation in which one may neglect the potential, and are therefore inconsistent with experiment.To construct a realistic $f(R)$ model requires at the very least a rather complicated function, with more than one adjustable parameter in order to fit the cosmological data and satisfy solar system bounds.

It is natural to consider generalizing such an action to include other curvature invariants~\cite{Carroll:2004de}, and it is straightforward to show these generically admit a maximally-symmetric solution: de Sitter space. Further, for a large number of such models (see e.g.~\cite{Navarro:2005gh}), solar system constraints, of the type I have described for $f(R)$ models, can be evaded. However, in
these cases another problem arises, namely that the extra degrees of freedom that arise are generically ghost-like.

An alternative, and particularly successful approach, is that employed by Dvali
and collaborators~\cite{Dvali:2000hr,Deffayet:2000uy,Deffayet:2001pu} in which an interesting modification to gravity arises from extra-dimensional models with
both five and four dimensional Einstein-Hilbert terms. These {\it Dvali-Gabadadze-Porrati (DGP) braneworlds} allow one to obtain cosmic acceleration from the gravitational sector because gravity deviates from the usual four-dimensional form at large distances. One may also ask whether ghosts plague these models. However, Dvali has claimed that this theory reaches the strong coupling regime before
a propagating ghost appears. In fact, Dvali has shown that theories that modify
gravity at cosmological distances must exhibit strong coupling phenomena, or else either possess ghosts or are ruled out by solar system constraints.

Current observational bounds are entirely consistent with a cosmological constant, but also with a range of dark energy models and the possibility that a modification to GR is the origin of cosmic acceleration. While it is often stated that one or other of these ideas is the simplest or most natural theoretical explanation, only increasingly accurate observations can settle the question and allow
us to make progress. In preparation for these, much theoretical work is necessary to extract concrete predictions with which to distinguish between the various
suggestions. A number of authors have already begin to tackle this problem, with one possible answer being that the cross-correlation of kinematical observables with tests involving the linear growth of structure as the universe expands~\cite{Jain03}. Whatever the ultimate answer, the accelerating universe looks bound
to teach us a deep truth about fundamental physics.

\vfill\eject

\section*{\centerline
{The Numerical Relativity Data Analysis Meeting
}}
\addtocontents{toc}{\protect\medskip}
\addtocontents{toc}{\bf Conference reports:}
\addcontentsline{toc}{subsubsection}{\it
Numerical Relativity-Data Analysis 
, by Patrick Brady}
\parskip=3pt
\begin{center}
Patrick Brady, University of Wisconsin, Milwaukee 
\htmladdnormallink{patrick-at-gravity.phys.uwm.edu}
{mailto:patrick@gravity.phys.uwm.edu}
\end{center}

The \emph{Numerical Relativity and Data Analysis} workshop that was
held at MIT on 6-7 November 2006 attracted 67 participants from both
the source modeling and data analysis communities. The meeting was
structured to encourage significant discussion by having only 4
speakers on the first day and 3 speakers on the second.  This meeting
had a rather narrow focus, dealing primarily with binary black holes;
the organizers hope that future meetings will address other important
sources.  Based on the hallway conversations, the meeting appears to
have succeeded in bringing together researchers from both communities. All
the talks, some rough notes from the discussions, and the list of
participants are posted on the meeting web site at
\htmladdnormallink
{\protect {\tt {http://www.lsc-group.phys.uwm.edu/events/nrda/}}}
{http://www.lsc-group.phys.uwm.edu/events/nrda/}

The meeting opened with a status report, by Ulrich Sperhake, on
numerical simulations of binary black holes.  The talk took a broad
view and reported on results from various groups, the technical status
of dynamical simulations, and touched on issues of initial data and
boundary conditions.  The discussion that followed included comments
by members of the numerical relativity community about the boundary
conditions, waveform extraction methods and radii, and convergence
testing to understand the accuracy of the simulations.  Data analysts
asked a number of questions about accuracy of the current simulations;
some of the numerical relativists turned the question around and asked
how accurate they need to be. These discussions continued through the
coffee break and led very naturally into the second talk.

Duncan Brown summarized the current status of searches for binary
black holes using data from gravitational-wave detectors.  In his
talk, he emphasized that mismatch--fractional loss of signal to
noise--is the correct measure of accuracy when discussing simulations
of gravitational waveforms for use in searches.  This point was
immediately picked up by those present; several groups had already
started to use the mismatch to understand the accuracy of their
simulations.  Brown further explained that sophisticated data analysis
pipelines are developed to deal with the non-Gaussian nature of the
gravitational-wave detector noise. He used this to emphasize that a
good match is necessary, but not sufficient in a matched filtering
search for signals. Finally, Brown pointed out that we should develop
a standard format for publishing data from the numerical relativity
community for use in searches for gravitational waves. Pablo Laguna
reported that there is already a collaboration (NRwaves) among
numerical relativists to collect their waveforms together for the sake
of making comparisons between the results.

The next presentation, by Mark Miller, addressed issues of numerical
accuracy in simulations. Mark invited the Caltech/Cornell and Jena
groups to present a summary of their investigations of accuracy in
numerical simulations. He followed that with a nice discussion of how
numerical relativity fits together with ongoing data analysis efforts.
In this discussion, he also presented a way to think about numerical
accuracy. His proposal generated considerable discussion among experts
in both numerical relativity and data analysis. Broadly speaking,
everybody agreed with trying to quantify the errors in the numerical
solutions, but precisely how to define the error remained unclear.

The last talk of the first day was given by Stephen Fairhurst. He
discussed the different sources of measurement error that affect
gravitational-wave observations. In particular, he emphasized the
difference between statistical and systematic (instrumental) errors if
the true waveform is accurately known. He explained how these issues
feed into current and future searches. In general, the required
accuracy of a simulation will depend on the accuracy with which the
instrumental response can be calibrated. Fairhurst finished by
explaining that this question is best answered by adding numerical
waveforms to real data and exploring our ability to detect and measure
them.

The first talk of the second day was given by Alessandra Buonanno. She
discussed comparisons between approximate analytically computed
waveforms and corresponding waveforms computed using numerical
relativity. For equal masses, she explained that both approaches (when
taken to sufficient accuracy) give very similar waveforms up to the
merger regime. It remains an open question 
to understand the physical origin of
the break in the numerically computed spectrum for these equal mass
systems and to explore the effects of spin on the waveforms.  Buonanno
finished by highlighting the need for numerical simulations that 
start from initial data that is physically close enough to a real
inspiral.

The workshop ended with Manuela Campanelli and Patrick Sutton
summarizing ``what we heard about ......'' data analysis and numerical
relativity, respectively.  Two points resonated through their
presentations and the following discussions.  First, making data from
numerical relativity simulations available for data analysis is highly
desirable, although some effort is needed to quantify the errors on
these data.  Second, this meeting was useful and people would like to
meet again to talk in more detail.
\vfill\eject
\section*{\centerline
{Note on the Numerical Relativity  
Data Analysis Meeting
}}
\addtocontents{toc}{\protect\medskip}
\addcontentsline{toc}{subsubsection}{\it
Note on Numerical Relativity-  
Data Analysis 
, by Peter Saulson}
\parskip=3pt
\begin{center}
Peter Saulson, University of Syracuse 
\htmladdnormallink{saulson-at-physics.syr.edu}
{mailto:saulson@physics.syr.edu}
\end{center}

As I sat in the back row of Rm NW14-1112 at MIT on Tuesday 7 Nov 2006, it 
suddenly struck me that we were participating in a watershed moment in the 
history
of gravitational physics. Here, in the same room, were two communities who 
decades earlier had promised to help each other in a grand adventure: the 
detection of gravitational waves and the use of those waves to explore the 
frontiers of strong field gravity. Then the difficulties of accomplishing grand things had intervened, and the years passed. But now, look what had been brought to the table. One the one hand, believable gravitational waveforms from multiple orbits of coalescing black hole binaries, checked and now cross-checked by a variety of independent methods and groups. On the other hand, operating interferometers at sites
around the globe, collecting data at sensitivities where detecting those black hole waveforms might come any day. (In their back pockets, plans for imminent upgrades of the interferometers that, when completed, might see one of those black
hole waveforms EVERY day.)

   Not only had each field suddenly reached a new level of maturity and accomplishment, but here were representatives from both sides struggling to understand each other's language in detail, so that the two communities could work effectively together. In real time, I saw numerical relativists adopt the data analysts'
``match'' parameter as an appropriate measure of error in their calculated waveforms, and gravity wave data analysts learn to read a waveform graph showing
${\Psi_4}(t)$ instead of $h(t)$.

Watershed moments don't happen often. I've spent twenty-five years working in gravitational wave detection, and I can't recall a scientific conference as transformative as this one. To find a parallel, I have to reach back to my earliest moments in the field, indeed to a meeting that happened before 
I had even started working in it. When I arrived as a green postdoc in 
Rai Weiss's MIT lab in the fall of 1981, Rai handed me a copy of the 
proceedings of the Batelle Seattle Workshop on Sources of Gravitational 
Radiation, which had been held in the summer of 1978. Reading this proceedings volume was the way that I was introduced to the state of the art of gravitational wave detection, and (at one remove) to many of the people in the field.

The summer of 1978 had been a crucial moment in the history of gravitational
wave detection. The community had weathered the controversy over Joe Weber's claims to have detected gravitational waves, and several groups were pressing forward with the new cryogenic bars. In the meantime, the nascent interferometer concept was only beginning to be seen as a possible way forward for the field. The introductory lecture in the Proceedings was a survey of detector technology by Rai Weiss, comparing bars and interferometers on the same basis and against different classes of signals. It clearly makes the bold claim that, if only the interferometer idea can be exploited at its natural scale of multi-kilometer arm length, then a dramatic step in sensitivity could be achieved, enough to make detection of gravitational waves likely.

The early days of numerical relativity are also recorded in several talks in
this volume. I'm not the best person to give a summary of those articles, but 
I'm willing to nominate these words of Larry Smarr as the most prophetic, ``This review has only scratched the surface of an immensely complicated subject. I hope
it will lead more people to think about these problems and give nonparticipants
some flavor of why progress sometimes seems so slow.''

One wonderful thing about these Proceedings is that the discussion sessions of the meeting were preserved in semi-verbatim format. The Discussion Session I:
Detection of Gravitational Radiation (transcribed by Reuben Epstein) is a gem. 
In it, you can see recorded, in real time, the realization that it was a good 
idea to push interferometers forward. Ron Drever makes the case clearly that 
interferometers ought to be funded alongside the already well-developed bars. 
When Steve Boughn raises a sensible caution about proceeding on too many 
fronts at once, the answers of Ron Drever, Dave Douglass, and Larry Smarr, 
representing the emerging consensus to move forward with interferometers, are 
carefully recorded, even down to Boughn’s undeserved put-down at the hands of 
Douglass, ``I think you will be more optimistic after you get your first 
cooldown.'' [laughter] There may have been a long interval yet before 
LIGO was actually born, but on these pages you can see the gleam in its 
parents' eyes.

Bob Forward gets the last word of the discussion in 1978, saying ``At least 
now we are able to draw the antenna sensitivity curves and the source 
[strength] curves on the same graph. Surely [laughter and applause] this 
means we have come
a long way.''

   In 1978, the field of gravitational wave detection was preparing to consolidate and move forward on two new fronts, interferometric detectors and numerical calculation of waveforms. In 2006, a standing-room-only crowd of scientists again learned to draw sensitivity curves and predicted waveforms on the same scales.
Only this time, the ultimate goal is finally in sight.
\vfill\eject

\section*{\centerline
{Unruh and Wald Fest}}
\addtocontents{toc}{\protect\medskip}
\addcontentsline{toc}{subsubsection}{
\it Unruh and Wald Fest, by
Carsten Gundlach and David Garfinkle}
\parskip=3pt
\begin{center}
Carsten Gundlach, University of Southampton
\htmladdnormallink{cg-at-maths.soton.ac.uk}
{mailto:cg@maths.soton.ac.uk}
\end{center}
\begin{center}
David Garfinkle, Oakland University 
\htmladdnormallink{garfinkl-at-oakland.edu}
{mailto:garfinkl@oakland.edu}
\end{center}

''A celebration of the careers and 60th birthdays of Bill Unruh and
Bob Wald'', held at the University of British Columbia, August 18-20, 2006

The meeting, with about 80 participants, had only four talks per day,
with plenty of time for pleasant interactions. The birthday
boys helped to pick their own speakers, and apparently felt they could be
selective!

Matt Choptuik spoke about ``The influence of Unruh and Wald on
numerical relativity'', a topic of personal interest to your
correspondents. Both had long been interested in cosmic
censorship. In 1991 Shapiro and Teukolsky famously claimed naked
singularity formulation in prolate collapse of collisionless matter
because singularities formed in the absence of an apparent horizon,
and Wald and Iyer then showed that even slices through Schwarzschild
need not have apparent horizons.

Choptuik then embarked on his own detailed study of spherical scalar
field collapse, which led to the discovery of critical phenomena in
gravitational collapse and a new, relatively ``natural'' way of
creating naked singularities. As a consequence, Hawking conceded his
bet with Preskill and Thorne that naked singularities could not form
from smooth initial data (for reasonable matter).

In 1993, Gregory and Laflamme conjectured that black strings might
become unstable and pinch off to form black holes; this would violate
cosmic censorship. Motivated by Horowitz through Unruh, Choptuik,
Lehner, Pretorius and Olabarrieta began investigating this numerically
in 2003, but the jury is still out.

In 1987 Thornburg wrote up a suggestion of Unruh's now known as
black hole excision (although he prefers singularity excision): no
boundary conditions are required on a boundary which can be spacelike
and stationary at once if it is inside a black hole. This was
implemented in 1992 by Seidel and Suen.

Abhay Ashtekar spoke on ``The quantum nature of the big bang''.
This was treated in loop quantum cosmology: a minisuperspace version of
loop quantum gravity in which one restricts to a small number of
degrees of freedom:
in this case Friedmann spacetime with
a scalar field so that the degrees of freedom are the scalar field
$\phi$ and the scale factor $a$. Instead of the Wheeler-de Witt equation
one obtains a difference equation for the wave function
$\psi(\phi,a)$. This peaks on a semiclassical trajectory which seems to go
through a bounce rather than to the big bang singularity.  Thus loop quantum
cosmology resolves the big bang singularity.  It will be interesting to
see whether this feature also holds in loop quantum gravity without
the minisuperspace approximation.

Jim Hartle spoke on ``What's wrong with your quantum mechanics?''
He imagined the objections that Bill and Bob, universally asknowledged as
deep thinkers and fierce critics, might have to the Gell-Mann-Hartle
consistent histories interpretation of quantum mechanics.  For each
objection, he then presented his response.  His
message was that {\it his} quantum
mechanics is as applicapable to the whole universe as it is to any
ordinary quantum mechanical system.  In each case, we divide the possible
outcomes for the system into ``approximately decohering coarse grained
histories'' and use the quantum state and the rules of quantum mechanics
to find the relative probability of each history.

Roger Penrose gave a public evening lecture ``What happened before the
big bang?'' He pointed out the puzzle of the very special initial conditions
of the universe and speculated that these might be connected to the final
conditions of a system that has radiated away all of its degrees of freedom.

Kip Thorne gave a review talk on ``Quantum non-demolition'' which was
both rich in historical detail, with much USA-USSR interaction and
competition, and quite technical. The field started with the
calculation by Braginsky in 1967 of the (standard) quantum limit for a
gravitational wave detector, motivated by the experimental work of
Weber. He stressed the crucial importance of two unpublished talks
Unruh gave in Bad Bentheim in 1981 which changed the focus from the
measurement of detector position to that of the classical force, which
can be measured to arbitrary accuracy. One current focus of the field
is to use LIGO to measure quantum mechanical effects on macroscopic
objects such as the mirrors.

Wojciech Zurek gave a review on the ``Emergence of the classical
world'' from quantum mechanics. He reviewed the Everett interpretation
in the light of the key questions ``what is the preferred basis?'' and
``why are probabilities the amplitude squared of those states?''.

Ted Jacobson gave a talk on ``Black hole entropy.'' Wald and
Parker showed in 1975 that Hawking radiation is in a thermal
state. Unruh showed that the key ingredient is the horizon. Even an
accelerated observer, who only has an ``acceleration horizon,'' sees
thermal radiation. Bekenstein proposed that black hole entropy counts the ways
in which a black hole could have formed. But how does thermodynamics
know about this?  Jacobson noted that the Hawking radiation has entropy
of its own and therefore must contribute something to the black hole
entropy.  He then considered the possibility that the entropy of the
radiation {\it is} in fact the entire black hole entropy.  A calculation
of the contribution of the radiation to the total entropy involves a
cutoff and so the answer seems to hinge on the appropriate value of the
cutoff to use.  At present, it is not clear what that appropriate value of
the cutoff is, so it is not clear whether the contribution of the Hawking
radiation to the black hole entropy is negligible, as had been assumed, or
dominant, as Jacobson seems to think likely.

Dick Bond gave a review talk on ``Inflation, gravitational waves and
the cosmic microwave background (CMB).'' Currently all observations
are compatible with cold dark matter and a simple cosmological
constant, and large-scale structure seeded by scale-invariant Gaussian
density fluctuations seeded by inflation. But much sophisticated
observation, analysis and modelling is behind this simple result, and
the limits on inflation history and the dark energy equation of state
are getting better. One new big goal is to see primordial
gravitational waves, through their interaction with CMB photons.

Ralf Sch\"utzhold
spoke on ``Effective horizons in the laboratory.''  He noted that though
black hole evaporation for stellar mass black holes is too small for us to hope
to measure it, there should be analogs of the Hawking effect that are within
reach of laboratory experiments.  These take place in optical systems and
in fluid systems where the medium is moving faster than the wave propagation
speed.  He presented the most promising such systems and for each system
the most promising
of the experimental techniques that might be used to detect the analog of the
Hawking effect in that system.

Stefan Hollands talked on ``Quantum fields in curved spacetime.''  
He emphasized that many of the ingredients used in specifying a quantum field theory in flat spacetime (spacetime symmetries, natural vacuum state, Euclidean methods, momentum
space, S matrix, etc.) are simply absent in curved spacetime.  One must therefore use completely different methods for quantum fields in curved spacetime, and Hollands and Wald advocate an algebraic approach that concentrates on the algebra of field operators and views quantum states as simply linear maps from the algebra to the complex numbers.  It has been known for some time how to do this for
free field theory; however it is only with the recent work of Hollands and Wald
that the groundwork has been laid for treating perturbative interacting quantum
fields in curved spacetime.  In particular, this work allows one to make sense of products of field operators.  These algebraic methods also allow one to formulate criteria for physically reasonable quantum states.

Gary Horowitz talked about ``Surprises in black hole evaporation.''  He 
noted that the standard picture of black hole evaporation within 
general relativity is that a black hole gives off thermal radiation 
until it reaches the Planck scale.  However, string theory takes place 
in more than 4 spacetime dimensions and involves extended objects. This 
gives rise to new possibilities.  There are higher dimensional analogs 
of black holes: black strings and black branes, which can be wrapped 
around extra spatial dimensions.  The horizon can then contain a 
topologically nontrivial circle.  Hawking radiation causes the size of 
this circle  to decrease.  When it becomes small enough, there is a 
“tachyon” instability.  This instability is due to certain modes of the 
string and causes  a  change in topology.  In the resulting spacetime  
the black hole is replaced by a ``bubble of nothing'' and simply 
disappears. This can occur when the curvature at the horizon is still 
small compared to the Planck scale.

\vfill\eject
\section*{\centerline
{Cliff Will Birthday Symposium }}
\addtocontents{toc}{\protect\medskip}
\addcontentsline{toc}{subsubsection}{\it
Will Fest, 
by Eric Poisson}
\parskip=3pt
\begin{center}
Eric Poisson, University of Guelph 
\htmladdnormallink{poisson-at-physics.uoguelph.ca}
{mailto:poisson@physics.uoguelph.ca}
\end{center}

{\bf Clifford Will is 60!} announced the webpage dedicated to this
Symposium, which was held in Saint Louis on Sunday, November 19th
2006, just a few days after Cliff's birthday (November 13th). The
Symposium came at the end of the 16th Midwest Relativity Meeting, an
always popular event during which all researchers, junior and senior,
contribute talks of 15 minutes. It says something about Cliff's
standing in the field that this was probably the best attended Midwest
Meeting in history, with over 90 participants. The Midwest Meeting was
organized by Wai-Mo Suen, Emanuele Berti, Jian Tao, Han Wang, and
Hui-Min Zhang from the Department of Physics at Washington University
in Saint Louis. The Symposium was organized by Wai-Mo Suen, Richard
Price, Bernard Schutz, Ed Seidel, S\'andor Kov\'acs, and Alan
Wiseman. The Symposium featured hour-long talks by invited speakers
Bernard Schutz, Luc Blanchet, Joseph Taylor, Francis Everitt, and Kip
Thorne. There was plenty of time between talks for coffee, discussion,
and poking fun at Cliff.

The first talk of the morning was by {\bf Bernard Schutz}, who gave
what he called the ``history talk,'' an overview of Cliff's career.
The talk was titled {\bf Will and Testament}, and it covered Cliff's
undergraduate-student days at McMaster University in Hamilton, Canada,
his graduate-student days at Caltech (working with Kip, of whom Cliff
had never heard before arriving at Caltech --- he only sought him out
on the advice of fellow Canadian graduate students), his
postdoc-days at the University of Chicago (working with Chandra), his
assistant-professor days at Stanford University (where he didn't get
tenure --- see below), and his distinguished career at Washington
University in Saint Louis, where Cliff is now James S.\ McDonnell
Professor of Physics. At the end of Bernie's talk, a member of the
audience asked whether Cliff had ever been known to be wrong on a
serious issue. Bernie answered that to his knowledge, this had never
happened. At this moment Leslie, Cliff's wife, raised an eager hand
and offered to present many examples of Cliff being in error. The
offer was declined, but Cliff explained: ``{\it My students are
frequently discouraged by the fact that, when we are in the middle of
some complicated post-Newtonian calculations and have a disagreement
over the coefficient of some term, I am almost always right. So I tell
them not to worry: at home, I've been right 67 times, while my wife
has been right 2,782,193 times.}''

The second talk of the morning was given by {\bf Luc Blanchet}, who
reviewed {\bf The Wonders of the Post-Newtonian}. This was a
fascinating talk during which Luc described the enormous progress that
has been accomplished in the last 15 years in the post-Newtonian
theory of two-body motion and gravitational-wave generation. This
effort has been pursued by a number of people around the world, with
Cliff and his collaborators playing an essential role. Among the
results obtained by these theorists is this impressive formula that
gives the rate at which a two-body system in circular motion loses
energy to gravitational radiation:
\begin{eqnarray*}
\frac{dE}{dt} &=& \frac{32c^5}{5G}\nu^2 x^5 \biggl\{ 1 +
  \left(-\frac{1247}{336} - \frac{35}{12}\nu \right) x + 4\pi x^{3/2} +
  \left(-\frac{44711}{9072} + \frac{9271}{504}\nu +
  \frac{65}{18} \nu^2\right) x^2
  \nonumber \\
  && \qquad \qquad \quad
  + \left(-\frac{8191}{672}-\frac{583}{24}\nu\right)\pi x^{5/2}
  \nonumber \\
  && \qquad \qquad \quad
  + \left[\frac{6643739519}{69854400}+
  \frac{16}{3}\pi^2-\frac{1712}{105}C -
  \frac{856}{105} \ln (16\,x) \right.
  \nonumber \\
  && \qquad \qquad \qquad ~
  + \left. \left(-\frac{134543}{7776} +
  \frac{41}{48}\pi^2
  \right)\nu - \frac{94403}{3024}\nu^2 -
  \frac{775}{324}\nu^3 \right] x^3
  \nonumber \\
  && \qquad \qquad \quad
  + \left(-\frac{16285}{504} +
  \frac{214745}{1728}\nu + \frac{193385}{3024}\nu^2\right)\pi x^{7/2} +
  {\cal O}\left(\frac{1}{c^8}\right) \biggr\}.
  \label{149}
\end{eqnarray*}
Here $x = (GM\omega/c^3)^{2/3}$ is a parameter (defined in terms of
the orbital frequency $\omega$ and the system's total mass
$M = m_1 + m_2$) that loosely represents $(v/c)^2$, the squared ratio
of orbital velocity to the speed of light, $\nu = m_1 m_2/M^2$ is a
dimensionless mass ratio, and $C \simeq 0.577$ is Euler's
constant. The hope is that the observational consequences of this
energy loss, which are manifested in the phasing of the gravitational
wave, will be verified by gravitational-wave detectors. This will
constitute a powerful test of general relativity, and as Luc pointed
out, an alternative way of measuring the mathematical constants $\pi$
and $C$.

The third and final talk of the morning was given by Nobel laureate
{\bf Joseph Taylor}. In his talk, titled {\bf Using and Testing
Relativity With Pulsars}, Joe reviewed the exciting history of binary
pulsars, which started in 1974 with his discovery (with then graduate
student Russell Hulse) of PSR 1916+13, and which has taken a recent
spectacular turn with the December 2003 discovery of the double pulsar
PSR J0737-3039. The handful of relativistic binary pulsars that have
been discovered to date have allowed sensitive tests of general
relativity to be performed, tests that probe strong-field and
radiative aspects of the theory. Nature could not have been more kind
to relativists! During his talk, Joe displayed the abstract page of
the first grant proposal in which he described plans for a systematic
search for radio pulsars; on this page appears a throw-away comment to
the effect that it would be a wonderful discovery if a pulsar could be
found within a binary system\ldots Joe also recalled the stimulating
discussions he had at Stanford, with Cliff and Bob Wagoner, on the
theoretical implications of his recent discovery.

The Symposium then broke for a group picture and lunch. (I went with
Cliff, Larry Kidder, and Patrick Brady to a nice place on Delmar
Boulevard. I had the chicken.) It resumed in the afternoon with
a talk by {\bf Francis Everitt}, titled {\bf Space, Gravity Probe B,
and Clifford Will}, in which he reviewed the long history of GPB, as
well as the exciting developments that followed its launch in April,
2004. The scientific goal of Gravity Probe B is to measure, for the
first time, the precession of test gyroscopes that is produced by the
gravity associated with Earth's rotational motion, thereby testing the
important relativistic prediction of frame dragging. Francis described
the effort that is now underway to analyze the terabyte of
experimental data that has been received from the probe to date. He
did not report results; for this we will have to wait until the April
2007 meeting of the American Physical Society. Francis also explained
Cliff's involvement in the project, mostly in his role as Chair of the
NASA Science Advisory Committee for Gravity Probe B.

The second talk of the afternoon was given by {\bf Kip Thorne}. In
{\bf Will and Waves}, Kip went a little deeper into historical matters
and recounted Cliff's scientific activities as a graduate
student. After Cliff spent some time talking with various researchers
at Caltech and JPL, he and Kip concluded that the time had arrived
(this was 1970) for a new generation of quantitative tests of general
relativity. Cliff started to think about a theoretical framework that
would facilitate the interpretation of the data, and would allow many
alternative theories to be contained within a unified package. In a
rapid burst of intense activity, he generalized the parameterized
post-Newtonian (PPN) framework that was introduced a few years earlier
by Ken Nordtvedt (building on earlier work by Eddington, Robertson,
Schiff, and others), and he proceeded to explore its consequences.
Cliff's version of the framework included a larger set of free
parameters, and it was based on a hydrodynamical description of the
matter instead of Nordtvedt's point-mass description.

In a period that started on August 24, 1970 and ended on May 1, 1972,
Cliff published 7 papers on this subject, a total of 105 pages in the
{\sl Astrophysical Journal}. (And Cliff got married to Leslie just two
months before! During his talk, Kip asked Cliff to describe his
honeymoon, but Cliff refused to comply.) In a first sequence of papers
(Theoretical Frameworks for Testing Relativistic Gravity I, II, and
III) he fleshed out the theoretical aspects of the PPN formalism. In
a second sequence of papers (Relativistic Gravity in the Solar System
I and II --- III was submitted when Cliff was a postdoc in Chicago) he
compared its predictions with astronomical data and placed bounds on
the free parameters. The mature form of the PPN framework, as it is
now displayed in Chapter 39 of Misner, Thorne, and Wheeler, was
presented in a third sequence of papers (Conservation Laws and
Preferred Frames in Relativistic Gravity I and II) co-authored with
Nordtvedt. Not bad for a mere graduate student!

Kip went on to describe the reasons why Cliff was not granted tenure
at Stanford, a topic that was alluded to by a number of speakers at
the Symposium. According to Kip, Stanford's standard for granting
tenure was that a candidate had to be one of the top three people
working in the field. Kip was asked to comment on Cliff's standing
within his peer group. As defined by Stanford, the peer group included
Roger Penrose, Stephen Hawking, and Kip Thorne himself\ldots Cliff was
not granted tenure, but Stanford's loss was WashU's gain.

The last word of the Symposium was left to the man himself:
{\bf Clifford Will}. During his {\bf Parting Shots}, Cliff
acknowledged the long list of people (colleagues, postdocs, students)
with whom he has collaborated and interacted in the course of his
career. He remarked that ``{\it what is so great about a career in
gravitational physics is the science and the people, rather than the
money or the power. I've noticed over almost 40 years in the business
that our field seems to have fewer than its share of arrogant,
mean-spirited, power-mad individuals, compared with other fields of
physics. I attribute this partly to the history of the field. For so
long general relativity was thought to be an irrelevant subject, in
the backwaters of physics and astronomy, so people who were full of
themselves, or out for the glory, would not find it attractive. Now
that gravitational physics has re-entered the mainstream of physics,
and has even taken on some of the characteristics of `big science,'
with things costing hundreds of millions, like Gravity Probe B and
gravitational-wave observatories, I hope that this will not change,
and that the field will continue to be populated by the kinds of
wonderful colleagues and friends I have encountered over my career.}''
Well said.

This concludes my description of the scientific component of the
Symposium. The event, however, included also a personal component, in
the form of a banquet for friends and family that took place on the
Saturday evening. (I had the chicken.) Cliff was paid a moving tribute
(in song) by the members of his family (daughters, sons-in-law, and
grandchildren) and was gently roasted by a group of his Saint-Louis
friends, who complained that he spends way too much time in Paris. He
was also (more vigorously) roasted by Alan Wiseman, who described
Cliff's tough-love approach toward the mentoring of graduate
students. (On a draft of an early research article written by Alan,
Cliff crossed out his own name, explaining that he did not want to be
associated with that piece of shit. I think he was kidding.)

The banquet's keynote act was a performance by {\bf Clifford and the
Silvertops}, a group of illustrious singers (also known as Bernie and
the Gravitones) consisting of Bernard Schutz (sporting a fake
mustache, singing lead, and playing the role of Clifford), Richard
Price, S\'andor Kov\'acs, and Kip Thorne (all with white-powdered
hair). To the tune of Paul Anka's My Way, they sang
\begin{verse}
{\large \bf Where there's a Will there's a way} \\

{\ } \\
It's {\bf been} a long time {\bf now} \\
I've {\bf work'd} with Einstein's {\bf theor}y \\
With {\bf work} and more work, {\bf wow} \\
No wonder {\bf why} I am so {\bf wear}y \\

{\ } \\
They {\bf asked} was Einstein {\bf wrong} \\
I told them {\bf no} and I earned {\bf high} pay \\
For {\bf math} so very {\bf long} \\
To do it {\bf his} way [The singers point at a lifesize picture of
Einstein.] \\

{\ } \\
New {\bf jobs} I've had to {\bf face} \\
But as to {\bf change} I now say {\bf foo}ey \\
I {\bf stay}, stay in one {\bf place} \\
I stay in {\bf France}, I mean Saint {\bf Loo}ey \\

{\ } \\
Geepee {\bf bee}, and geepee {\bf ess} \\
Gee, whiz I {\bf guess} that we can {\bf now} say \\
Na{\bf ture} has passed the {\bf test} \\
She did it {\bf his} way \\

{\ } \\
When I was {\bf young}, Newton was {\bf all} \\
But then came {\bf post}, and that's not {\bf all} \\
After the {\bf post}, a host more {\bf post} \\
Until I {\bf thought} that I was {\bf toast} \\
A billion {\bf terms}, a can of {\bf worms} \\
To do it {\bf his} way \\

{\ } \\
Up {\bf north} people are {\bf few} \\
We almost {\bf never} [5 silent beats] spoke \\
But {\bf here} to be a {\bf jew} \\
They made me {\bf learn} to tell a {\bf bad} joke \\

{\ } \\
I've {\bf friends}, I think I {\bf do} \\
And colleagues {\bf some} who made my {\bf hair} grey \\
So {\bf long}, so long a{\bf go} \\
Doing it {\bf his} way \\

{\ } \\
I ruled the {\bf field}, and here's the {\bf thing} \\
My work, my {\bf book}, I was a {\bf king} \\
I was the {\bf star} where'er I'd {\bf roam} \\
But time to {\bf time} I would come {\bf home} \\
Home to my {\bf life}, home to my {\bf wife} \\
To do it {\bf her} way [The singers point at Leslie.]

\end{verse}
The lyrics to this great song were written by Richard Price, and they
are reproduced here with his kind permission. The bold words are
emphasized (held longer) to keep beat with the music.

I'll close this report with a personal note. I have a vivid memory of
the time when Cliff offered for me to come to Saint Louis and work with
him as a postdoc. I was overjoyed! After my time at Caltech this was
where I most wanted to be. It has been my great fortune and privilege
to work with Cliff, and I am proud to count him as a friend. I am very
glad to have been a participant at this Symposium, and I wish Cliff a
very happy 60th birthday.

[I thank Richard Price for his permission to reproduce the song's
lyrics, and Clifford Will for providing me with the italicized
quotes. I thank them both for fact-checking an earlier draft of this
report and providing suggestions for improvement.]

\vfill\eject

\section*{\centerline
{Brane-World Gravity: Progress and Problems
}}
\addtocontents{toc}{\protect\medskip}
\addcontentsline{toc}{subsubsection}{\it
Brane-World Gravity
, by Andrew Mennim}
\parskip=3pt
\begin{center}
Andrew Mennim, University of Portsmouth 
\htmladdnormallink{Andrew.Mennim-at-port.ac.uk}
{mailto:Andrew.Mennim@port.ac.uk}
\end{center}

The Institute of Cosmology and Gravitation hosted a two-week international conference at the end of September on the subject of brane-world gravity.  The conference began with a three-day meeting which was followed by a workshop; 
about 80 delegates attended.  The programme and slides from most of the 
talks can be found on the conference website, the URL for which is 
\htmladdnormallink
{\protect {\tt {http://www.icg.port.ac.uk/brane06/}}}
{http://www.icg.port.ac.uk/brane06/}

Invited speakers
were Cliff Burgess, Cedric Deffayet, Gary Gibbons, Ruth Gregory, 
Panagiota Kanti, David Langlois, James Lidsey, Kei-ichi Maeda, Nick Mavromatos, Lefteris Papantanopoulos, Valery Rubakov, Misao Sasaki, Tetsuya Shiromizu, Jiro Soda, Kellogg Stelle and Takahiro Tanaka.

Brane-world models have been studied intensively for the last decade.  Originally motivated by the existence of branes in string theory, brane-worlds have been
of interest to the particle physics community because they offer new ways to explain hierarchies, and because of the new phenomenology for colliders and cosmic
ray showers resulting from the possibility of a low Planck mass.  They have also inspired relativists and cosmologists because they represent a very geometrical way to modify gravity and to change the cosmological history of the universe.
The conference focussed on the gravitational and cosmological aspects of brane-worlds, the aim being to review recent progress in the field and to spark discussions and collaborations on
the outstanding issues.

The themes discussed in the meeting were cosmology and the evolution of cosmological perturbations in brane-worlds, the  Dvali--Gabadadze--Porrati (DGP) model and its possible problems with ghosts, the nature of black holes in brane-worlds
and possible collider signatures, possible solutions to the cosmological constant problem using six-dimensional brane-worlds, and links between the phenomenological models and fundamental physics ideas like string theory.  The meeting ended with a discussion of the outstanding issues, identifying projects for study during the workshop and beyond.
About half of the delegates remained for the workshop.  The workshop involved two talks each day with time in between for delegates to discuss the themes raised and form collaborations.

Some interesting subjects and outstanding questions were discussed, resulting in an advance in understanding and new collaborations.  Effective actions are very useful tools in higher-dimensional physics, but it is important to understand in which circumstances they are effective;  for Kaluza--Klein theories this is entirely understood but for non-homogeneous configurations there are additional subtleties.  Understanding the quantum vacuum state for the early universe in the
Randall--Sundrum model with inflation on the brane is important for predicting possible cosmological signatures;  it was argued by some that the initial state could be and by others that it must be very close to the usual four-dimensional result.
Perhaps most contentious was the issue of ghost states in the DGP model.  Some delegates presented work showing that the model has a ghost state either in the spin-two or spin-zero sector, but it was argued by others that this does not necessarily invalidate the model because the energy scale associated is on the limit of where one can trust an effective four-dimensional description.

The local organising committee (Kazuya Koyama, Andrew Mennim and Sanjeev Seahra) would like the thank David Langlois, Roy Maartens, Kei-ichi Maeda, Lefteris Papantanopoulos, Misao Sasaki and David Wands for their help in the organisation of the conference; and the Institute of Physics, and the Particle Physics and Astronomy Research Council for providing financial support.

\vfill\eject

\section*{\centerline
{Workshop on Gravity and Theoretical Physics 
}}
\addtocontents{toc}{\protect\medskip}
\addcontentsline{toc}{subsubsection}{
\it Gravity and Theoretical Physics 
, by Marco Cavaglia}
\parskip=3pt
\begin{center}
Marco Cavaglia, University of Mississippi 
\htmladdnormallink{cavaglia-at-phy.olemiss.edu}
{mailto:cavaglia@phy.olemiss.edu}
\end{center}

The second School \& Workshop on Gravity and Theoretical Physics was held
at the main campus of The University of Mississippi (Oxford, MS) on
January 8-11, 2007.

The purpose of this meeting, the second in a series, was to to bring
together researchers and graduate students to exchange ideas on field
theory, gravity and related areas. The program consisted of a series of
lectures from faculty and senior researchers and shorter talks by graduate
students, who were given the opportunity to present their current
research. About 20 participants from Mississippi, Alabama and Kentucky
attended the workshop.  No registration fee was charged and all talks were
open to the public. A visit to William Faulkner's Rowan Oak and the local
University museum were part of the social activities. This event was
possible thanks to the generous support of the Department of Physics and
Astronomy of The University of Mississippi.

School lectures were given by Keith Andrew and Brett Bolen (Western
Kentucky U.), Luca Bombelli, Vitor Cardoso, Marco Cavaglia and Itai Seggev
(U. Mississippi), Lior Burko (U. Alabama - Huntsville) and Ben Harms (U.
Alabama - Tuscaloosa). Main topics included finite temperature field
theory, statistical geometry, gravitational waves and black hole physics.
Student talks covered a wide range of issues, ranging from non-commutative
physics to extra-dimensional models and gravitational radiation. The
program and the presentations delivered in digital format are available at
the school webpage 
\htmladdnormallink
{\protect {\tt {http://www.phy.olemiss.edu/GR/gravity07/}}}
{http://www.phy.olemiss.edu/GR/gravity07/}

The University of Mississippi plans to make this an annual event, with the
next school tentatively scheduled for January 2008. For further
information, please contact Marco Cavaglia (cavaglia@olemiss.edu).

\end{document}